\documentstyle[aps,epsfig,multicol,eqsecnum,psfrag]{revtex}    

\newcommand{\mycaption}[2]{\begin{center}{\bf Figure \thefigure}\\{#1}\\{\em #2}\end{center}\addtocounter{figure}{1}}
\newcommand{\myauthor}{Simonsen, Leskova, and Maradudin}
\newcommand{\mytitle}{Light scattering from an amplifying medium bounded by a randomly
  rough surface:\hspace*{0.1cm} A numerical study}

\begin{document}

\title{\mytitle}

\author{Ingve Simonsen$^{a,b}$, Tamara A.\ Leskova$^c$, Alexei A.\  Maradudin$^b$,}
\address{$^a$Department of Physics, The Norwegian University of 
Science and Technology,  Trondheim, Norway}
\address{$^b$Department of Physics and Astronomy and Institute for Surface and
Interface Science,\\
University of California, Irvine, CA 92697, USA}
\address{$^c$Institute of Spectroscopy, Russian Academy of Sciences,
Troitsk 142092, Russia}

\date{May 7, 2000}
\maketitle

\begin{abstract}
  We study by numerical simulations the scattering of $s$-polarized
  light from a rough dielectric film deposited on the planar surface of  a
  semi-infinite perfect conductor. The dielectric film is allowed to
  be either active or passive,  situations that we model
  by assigning  negative and positive values, respectively, to 
  the imaginary part $\varepsilon_2$ of the
  dielectric constant of the film. We study the
  reflectance ${\cal R}$ and the total scattered energy ${\cal U}$ for
  the system as functions of both $\varepsilon_2$ and the angle of
  incidence of the light. Furthermore, the positions and widths of the
  enhanced backscattering and satellite peaks are discussed. It is
  found that these peaks become narrower and higher when the
  amplification of the system is increased, and that their widths
  scale linearly with $\varepsilon_2$.  The positions of the
  backscattering peaks are found to be independent of $\varepsilon_2$,
  while we find a weak dependence on this quantity in the positions of
  the satellite peaks.  \date{\today}
\end{abstract}
\pacs{PACS numbers: }


\section{Introduction}

In the first half of the 1990's and subsequently, amplifying volume
disordered media received a great deal of attention from
theorists~\cite{Freilikher,Review} and
experimentalists~\cite{Review,Wiersma} alike. This attention was
partly motivated by the suggestion of using random volume scattering
media to construct a so-called random laser~\cite{RandomLaser}.  For
scattering systems possessing surface disorder in contrast to volume
disorder, the overwhelming majority of theoretical and experimental
studies have been devoted to scattering from passive ({\it i.e.}
absorbing) media.  Only recently has the surface scattering community
begun on studies of surface disordered amplifying systems. The only
literature on the scattering of light from amplifying surface
disordered media known to us is the theoretical study by Tutov {\em et
  al.}~\cite{Tutov} and the experimental investigation by Gu and
Peng~\cite{Gu}. In the theoretical work by Tutov {\it et
  al.}~\cite{Tutov} the authors conducted a perturbative study of the
scattering of $s$-polarized light from an amplifying film deposited on
the planar surface of a perfect conductor, where the vacuum-film
interface was a one--dimensional random interface characterized by a
Gaussian power spectrum.  In this work we consider the same scattering
system, but apply a numerical simulation approach for its study. The
numerical approach is based on the solution of the reduced Rayleigh
equation that the scattering amplitude for the system satisfies.  The
use of a numerical simulation approach enables us to study possible
non-perturbative effects~\cite{Simonsen99} that could not be accounted
for by the perturbative technique used in Ref.~\cite{Tutov}.
Furthermore, we also use a different power spectrum of the surface
roughness. In particular, a West-O'Donnell (or rectangular) power
spectrum~\cite{west95} is used in this work, in contrast to the
Gaussian power spectrum used by Tutov~{\em et al.} Such a power
spectrum allows for the suppression of single scattering over a range
of scattering angles and, more important, it opens the possibility for
a strong coupling of the incident light to guided waves supported by
the film structure.

In this work we calculate the reflectivity when a system consisting of
vacuum in the region $x_3 > \zeta (x_1)$, an amplifying medium in the
region $-d < x_3 < \zeta (x_1)$, and a perfect conductor in the region
$x_3 < -d$, is illuminated from the vacuum side by s-polarized light
of frequency $\omega$. The surface profile function $\zeta (x_1)$ is
assumed to be a single-valued function of $x_1$ that is differentiable
as many times as is necessary, and  constitutes a zero-mean,
stationary Gaussian random process characterized by a West--O'Donnell height
autocorrelation function.  The amplifying medium is modeled by a
dielectric medium whose dielectric constant $\varepsilon$ has an
imaginary part $\varepsilon_2$ that is negative, while its real part
$\varepsilon_1$ is positive.  The values of $\varepsilon_2$ are chosen
so that they include gains
($g=2\pi|\varepsilon_2|/(\lambda\sqrt{\varepsilon_1})$) in the medium
that are physically realizable.  The assumption of a negative
imaginary part to $\varepsilon$ is the simplest way of modeling
stimulated emission in this system.  The reflectivity is given by
$|R(k)|^2$, where $R(k)$ is defined in terms of the scattering
amplitude $R(q|k)$ by $|\langle R(q|k)\rangle |^2$ $= L_1 2\pi\delta
(q-k)|R(k)|^2$.  In this relation the wavenumbers $k$ and $q$ are
related to the angles of incidence and scattering by $k = (\omega
/c)\sin\theta_0$ and $q = (\omega /c)\sin\theta_s$, respectively,
$L_1$ is the length of the $x_1$-axis covered by the random surface,
and the angle brackets denote an average over the ensemble of
realizations of the surface profile function $\zeta (x_1)$.  The
scattering amplitude $R(q|k)$ is obtained by solving numerically the
reduced Rayleigh equation it satisfies for a large number of
realizations of $\zeta (x_1)$, and $\langle R(q|k)\rangle$ is obtained
by averaging the results.  As expected, the reflectivity of the
amplifying medium with a random surface is larger than that of the
corresponding absorbing medium, viz. a medium with the same value of
$|\varepsilon_2| $ but with $\varepsilon_2$ positive, for all angles
of incidence.

\section{Scattering Theory}

\subsection{Scattering system}

The scattering system that will be considered in this paper consists
of a dielectric film, with a randomly rough top interface, deposited
on the planar surface of a semi-infinite perfect conductor. In
particular, it consists of vacuum in the region $x_3>\zeta(x_1)$, an
amplifying or absorbing dielectric medium in the region
$-d<x_3<\zeta(x_1)$, and a perfect conductor in the region $x_3<-d$. This
geometry is depicted in Fig.~1. The rough surface profile
function, denoted by $\zeta(x_1)$, is assumed to be a single-valued
function of its argument, and differentiable as many times as
needed. Furthermore, it is assumed to constitute a zero-mean,
stationary, Gaussian random process defined by
\begin{mathletters}
  \begin{eqnarray}
    \left< \zeta(x_1)\right> &=& 0,\\
    \left< \zeta(x_1) \zeta(x_1')\right> &=& \delta^2 W(|x_1-x_1'|),
  \end{eqnarray}
\end{mathletters}
where $<\cdot>$ denotes an average over the ensemble of realizations
of $\zeta(x_1)$, and $\delta$ is the rms-height of the rough surface.
Moreover, $W(|x_1|)$ denotes the surface height autocorrelation
function, and is related to the power spectrum of the surface
roughness $g(|k|)$ by
\begin{eqnarray}
    \label{psd}
    g(|k|)  &=& \int^{\infty}_{-\infty} dx_1  W(|x_1|)\;  e^{-ikx_1}.
\end{eqnarray}
In the numerical simulation results to be presented later, we will 
assume a rectangular power spectrum, also known as the West-O'Donnell form, 
\begin{eqnarray}
    \label{psd-west-o'Donnell}
    g(|k|)  &=& \frac{\pi}{k_{+}-k_{-}}
    \left[  \theta(k-k_{-}) \theta(k_{+}-k)
          + \theta(-k_{-}-k) \theta(k+k_{+})\right] ,
\end{eqnarray}
where $\theta (k)$ is the Heaviside unit step function, and $k_{\pm}$
are parameters to be specified. This power spectrum has recently been
used in an experimental study of enhanced backscattering from weakly
rough surfaces~\cite{west95}.  

\vspace*{1cm}
\begin{figure}[b!]
    \begin{center}
          \epsfig{file=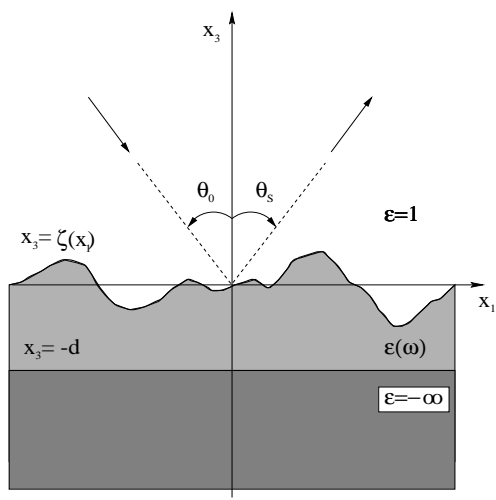, width=6.5cm,height=6.5cm} 
    \end{center}
    \caption{The scattering geometry considered in the present
      work.}
   \label{fig1}
\end{figure}


\subsection{Scattering Equations}

If the vacuum-dielectric interface $x_3=\zeta(x_1)$ is illuminated
from the vacuum side by an s-polarized electromagnetic wave of
frequency $\omega$, the only nonzero component of the electric field
vector in the region $x_3 > \zeta(x_1)_{max}$ is the sum of an
incident wave and a scattered field:
\begin{eqnarray}
  E_2^>(x_1,x_3|\omega) &=& 
  e^{ikx_1 - i\alpha_0(k,\omega) x_3} 
      + \int^{\infty}_{-\infty} \frac{dq}{2\pi} 
           R(q|k)e^{iqx_1+i\alpha_0(q,\omega) x_3 }.
\end{eqnarray}
In this equation $R(q|k)$ denotes the scattering amplitude, while we
have defined 
\begin{eqnarray}
  \alpha_0(q,\omega) &=& 
  \left\{ 
    \begin{array}{cl}
      \sqrt{\frac{\omega^2}{c^2}-q^2}, \quad & |q| < \omega /c\\
      i\sqrt{q^2 - \frac{\omega^2}{c^2}}, \quad & |q| > \omega /c.
    \end{array}
  \right.
\end{eqnarray}

From a knowledge of the scattering amplitude one can define the
differential reflection coefficient~(DRC) $\partial
R/\partial\theta_s$. It is defined such that $(\partial R/\partial
\theta_s) d\theta_s$ is the fraction of the total time-averaged flux
incident on the surface that is scattered into the angular interval
$d\theta_s$ about the scattering angle $\theta_s$, in the limit as
$d\theta_s \rightarrow 0$.  The contribution to the mean differential
reflection coefficient from the coherent (specular) component of the
scattered field is given by~\cite{AnnPhys,PhysRep}
\begin{mathletters}
    \label{Eq:DRC}
\begin{eqnarray}
    \label{Eq:DRC-a}
  \left< \frac{\partial R}{\partial\theta_s}\right>_{\rm coh} &=& 
  \frac{1}{L_1} \frac{\omega}{2\pi c}
  \frac{\cos^2\theta_s}{\cos\theta_0} 
     \left|\left< R(q|k)\right> \right|^2, 
\end{eqnarray}
and the contribution to the mean differential reflection coefficient
from the incoherent (diffuse) component of the scattered field is
given by~\cite{AnnPhys,PhysRep}
\begin{eqnarray}
    \label{Eq:DRC-b}
  \left< \frac{\partial R}{\partial \theta_s}\right>_{\rm incoh} &=& 
  \frac{1}{L_1}\frac{\omega}{2\pi c}
  \frac{\cos^2\theta_s}{\cos\theta_0} 
  \left[
    \left< \left|R(q|k)\right|^2\right> 
       -\left|\left< R(q|k)\right>\right|^2
   \right] .
\end{eqnarray}
\end{mathletters}
In Eqs.~(\ref{Eq:DRC}), $L_1$ is the length of the $x_1$-axis covered
by the random surface, and the wave numbers $k$ and $q$ are
related to the  angles of incidence $\theta_0$
and the angle of scattering $\theta_s$ 
according to
\begin{eqnarray}
  \label{Eq:momentum}
  k &=& \frac{\omega}{c}\sin\theta_0 , \qquad 
  q = \frac{\omega}{c}\sin\theta_s .
\end{eqnarray}
Both these angles are measured from the normal to the mean surface
as indicated in Fig.~1.

From the definition of the mean differential reflection coefficient,
we find that the reflectance of the surface is defined according to
\begin{eqnarray}
  \label{Eq:reflectance}
  {\cal R} &=& \int^{\frac{\pi}{2}}_{-\frac{\pi}{2}} d\theta_s\,
          \left< \frac{\partial R}{\partial\theta_s}\right>_{\rm coh}
          \;=\;
          \left| R\left( k \right)\right|^2 ,  
\end{eqnarray}
where $k$ is given by Eq.~(\ref{Eq:momentum}), and
$R(k)$ is related to the scattering amplitude $|R(q|k)|^2$ by 
$|\left<R(q|k)\right>|^2=L_1 2\pi\delta(q-k)\left|R(k)\right|^2$.
Likewise, the total scattered energy (normalized to the incident
energy) is defined by   
\begin{eqnarray}
  \label{Eq:total-energy}
  {\cal U} &=& \int^{\frac{\pi}{2}}_{-\frac{\pi}{2}} d\theta_s\,
          \left< \frac{\partial R}{\partial\theta_s}\right>,  
\end{eqnarray}
where $\left< \partial R/\partial\theta_s\right>$ is the total mean
DRC, {\it i.e.} the sum of the coherent and incoherent contribution as
defined in Eqs.~(\ref{Eq:DRC-a}) and (\ref{Eq:DRC-b}), respectively.

So far we have not specified how to obtain the scattering amplitude
entering into the above equations. It has previously been shown that
$R(q|k)$ is the solution of the so-called reduced Rayleigh
equation~\cite{RRE}. This single, inhomogeneous integral equation for
$R(q|k)$ for our scattering geometry reads~\cite{PhysRep} 
\begin{mathletters}
  \label{Eq:RRE}
\begin{eqnarray}
   \int^{\infty}_{-\infty} \frac{dq}{2\pi} M(p|q)R(q|k) &=& N(p|k) ,
\end{eqnarray}
where
\begin{eqnarray}
  M(p|q) &=& 
  \frac{e^{i\alpha(p,\omega) d}}{\alpha_0(q,\omega)+\alpha(p,\omega)} 
         I(\alpha_0(q,\omega) + \alpha(p,\omega) |p-q) 
     \nonumber \\ && \mbox{} \quad 
  - \frac{e^{-i\alpha(p,\omega)
  d}}{\alpha_0(q,\omega)-\alpha(p,\omega)} 
         I(\alpha_0(q,\omega) - \alpha(p,\omega) |p-q)\\*[2mm] 
  N(p|k) &=& 
  -\frac{e^{i\alpha(p,\omega)d}}{\alpha(p,\omega)-\alpha_0(k,\omega)} 
        I(\alpha(p,\omega) - \alpha_0(k,\omega) |p-k)
     \nonumber \\ && \mbox{} \quad
   - \frac{e^{-i\alpha(p,\omega)d}}{\alpha(p,\omega)+\alpha_0(k,\omega)} 
     I(-\alpha(p,\omega) - \alpha_0(k,\omega) |p-k),
\end{eqnarray}
with 
\begin{eqnarray}
  I(\gamma |q) = \int^{\infty}_{-\infty} dx_1 e^{i\gamma\zeta(x_1)} e^{-iqx_1} .
\end{eqnarray}
\end{mathletters}
In writing Eq.~(\ref{Eq:RRE}) we have introduced  
 \begin{eqnarray}
   \alpha(q,\omega) &=& \sqrt{\varepsilon(\omega)
  \frac{\omega^2}{c^2} - q^2},
\end{eqnarray}
where the branch of the square root is chosen so that the real part of
$\alpha(q,\omega)$ is always positive while the imaginary part is
positive when $\varepsilon_2 > 0$, but is negative when $\varepsilon_2
< 0$.

The simulation results to be presented in the next section were
obtained by directly solving numerically the reduced Rayleigh
equation~(\ref{Eq:RRE}).  This nonperturbative approach can treat 
much longer rough
surfaces as compared to a rigorous numerical simulation
approach~\cite{AnnPhys} with the same use of computer power and
memory. An additional advantage of a numerical approach based on the
reduced Rayleigh equation is that  ${\cal R}$ and ${\cal U}$ can be
calculated to high precision, whereas the same quantities calculated
by a rigorous approach have been found to be less accurate for the
surface lengths typically used in such simulations.  This difference
in accuracy for ${\cal R}$ and ${\cal U}$ for these two numerical
approaches we believe is related to the difference in the length of
the surface that can be handled practically with today's typical
computer resources.  The numerical solution of the reduced Rayleigh
equation is done by converting the integral equation into a set of
linear equations obtained by using an appropriate quadrature scheme
and solving the resulting system by standard numerical
techniques~\cite{NR}.  Due to increased numerical performance, the
calculation of the $I(\gamma |q)$-integrals was based on an expansion
of the integrand in powers of the surface profile function.  This
numerical method has recently been applied successfully to a similar
scattering geometry~\cite{Simonsen99}, and the interested reader is
directed to this paper for details of the numerical method.

\section{Results and discussions}

For the numerical simulations to be presented below we have considered
the scattering of $s$-polarized incident light of wavelength
$\lambda=632.8\,{\rm nm}$.  The film was assumed to have mean
thickness $d=500\,{\rm nm}$, and its dielectric constant at the
wavelength of the incident light was taken to be $\varepsilon(\omega)
= 2.6896+i\varepsilon_2$, where $\varepsilon_2$ is allowed to vary
over both positive and negative values. The surface profile function
was characterized by a
power spectrum  of the West-O'Donnell type as defined in
Eq.~(\ref{psd-west-o'Donnell}). For the parameters defining the power
spectrum we used $k_-=0.86\, \omega/c$ and $k_+= 1.97\, \omega/c$.
For these values of $k_\pm$, single scattering should be suppressed
for scattering angles in the range $|\theta_s|< 55.1^\circ$.
The rms-height of the surface was taken to be $\delta=30 {\rm nm}$.
Furthermore, the length of the surface was taken to be $L=160\lambda$,
and the numerical results were all averaged over $N_\zeta=3000$
realizations of the surface profile function.

In Fig.~2a we present the numerical simulation results for
the contribution to the mean differential reflection coefficient from
the light that has been scattered incoherently, $\left<\partial
  R/\partial \theta_s\right>_{{\rm incoh}}$, for $s$-polarized light
incident normally on the mean surface ($\theta_0=0^\circ$).  The
values of the imaginary part of the dielectric function were (from
top to bottom) $\varepsilon_2= -0.0025$, $0$, and $0.0025$.  From this
figure we notice the enhanced backscattering peaks located at
$\theta_s=\theta_0=0^\circ$.  Moreover, two satellite peaks, located
symmetrically about the position of the enhanced backscattering peak,
are easily distinguished from the background. Their positions as read
of from Fig.~2a fit nicely with their positions, $\theta_\pm
= \pm 17.7^\circ$, calculated for the corresponding planar geometry in
the limit of vanishing $\varepsilon_2$~\cite{Simonsen99}. The choices
made for $\varepsilon_2$ of the film in Fig.~2a correspond
to an amplifying or active film ($\varepsilon_2=-0.0025$), a neither
amplifying nor absorbing film ($\varepsilon_2=0$), and an absorbing or
passive film ($\varepsilon_2=0.0025$), respectively.  This is
reflected in Fig.~2a where the contribution to the mean DRC
from the light scattered incoherently from the amplifying medium is
larger for all scattering angles then for the other two cases due to
the extra energy gained by the scattered light from the amplifying
film. Moreover, it is interesting to notice that the differences
between these curves are largest for small scattering angles, and 
as one moves to larger scattering angles these differences are
reduced.  The main reason for this is that for scattering angles
$|\theta_s|<55.1^\circ$ the light has undergone multiple scattering
which will increase the effects of amplification and absorption as
compared to a system that is dominated by single scattering events for
such scattering angles. In the wings of the angular dependence of the
mean DRC, $|\theta_s|>55.1^\circ$, where single scattering gives the
main contribution, the differences between the curves
corresponding to different values of $\varepsilon_2$ is much less
pronounced.  In order to obtain a more complete picture of how the
incoherent component of the mean DRC depends on the imaginary part of
the dielectric function, in Fig.~2b we present a plot
showing $\left<\partial R /\partial \theta_s\right>_{{\rm incoh}}$ as
function of $\varepsilon_2$, as well as of the scattering angle
$\theta_s$. The angle of incidence here was also chosen to be
$\theta_0=0^\circ$. As seen from this plot, the positions of the peaks
are fixed, or close to fixed, while the overall amplitude of
$\left<\partial R /\partial \theta_s\right>_{{\rm incoh}}$ increases
monotonically with decreasing imaginary part of the dielectric
constant.


\vspace*{0.8cm}
\begin{figure}[b!]
  \begin{center}
    \epsfig{file=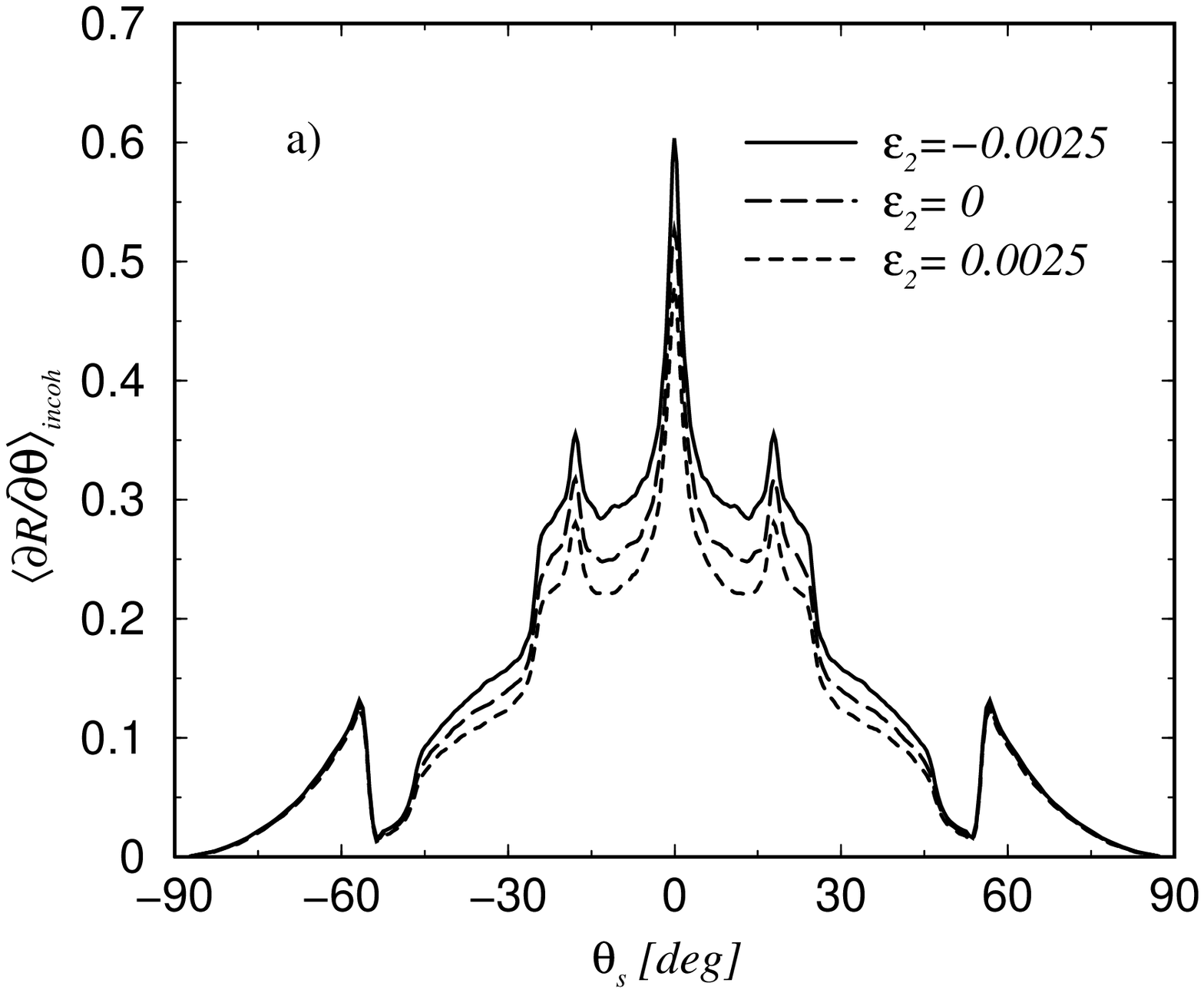, width=7.5cm,height=7.2cm} \\*[0.3cm]
    \psfrag{x}{$\theta_s$} \psfrag{y}{$\varepsilon_2$}
    \psfrag{z}{$\left<\frac{\partial R}{\partial \theta_s}\right>_{\rm
        incoh}$}
    \epsfig{file=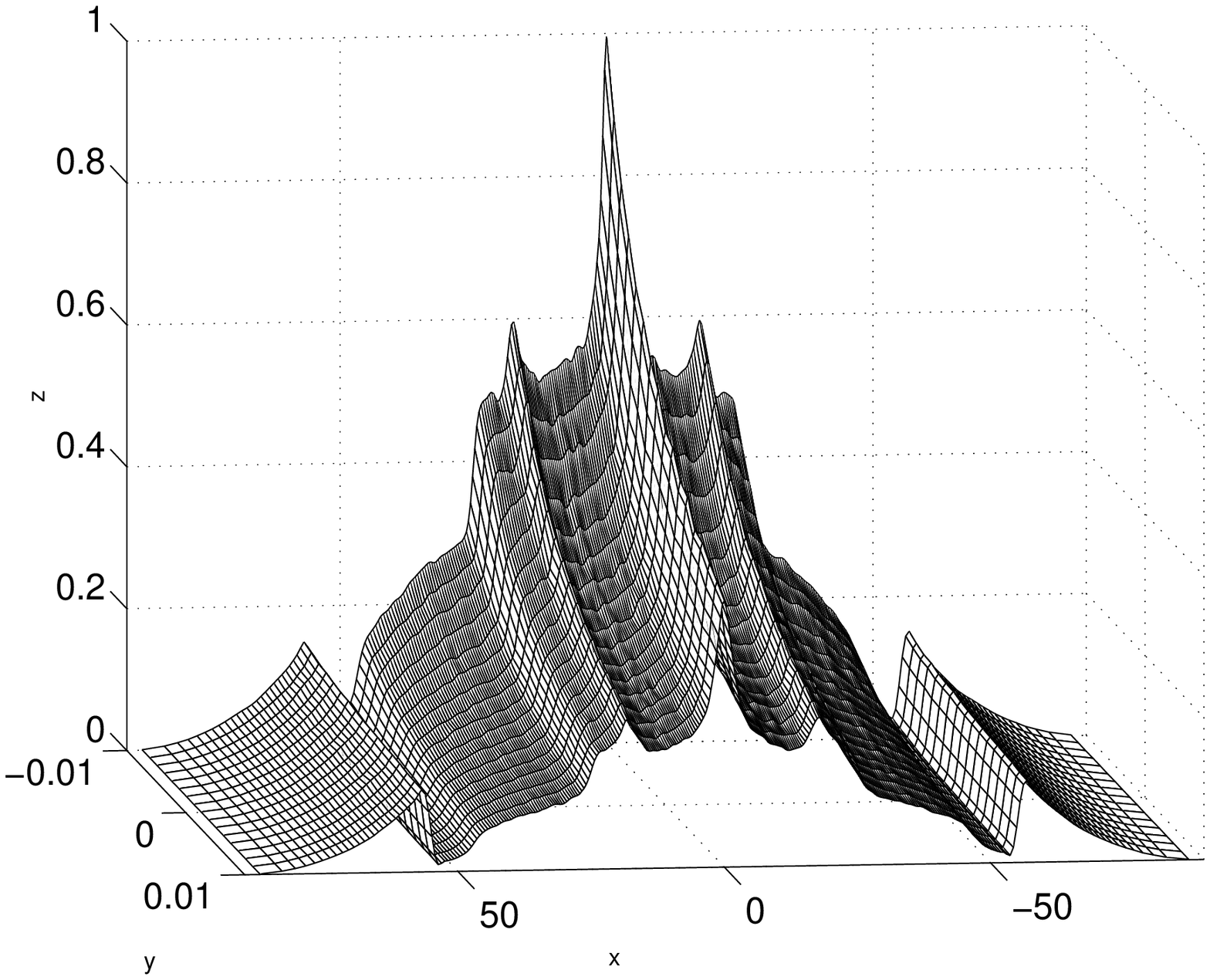,width=8.5cm,height=7.2cm}
  \end{center}
  \caption{The mean differential reflection coefficient for the
  incoherently scattered light for (a) $\varepsilon_2=0,\pm 0.0025$
  and (b) as a function of the same parameter. For both figures the
  angle of incidence was $\theta_0=0^\circ$, and the wavelength of the
  incident light was $\lambda=632.8\,{\rm nm}$.  The dielectric
  function of the film of mean thickness $d=500 {\rm nm}$ was
  $\varepsilon(\omega) = 2.6896+i\varepsilon_2$, where $\varepsilon_2$
  is as indicated in the figure. The randomly rough surface had an rms
  roughness of $\delta=30 {\rm nm}$.  The power spectrum was of the
  West-O'Donnell type defined by the parameters $k_-=0.86\, \omega/c$
  and $k_+= 1.97\, \omega/c$.}
\label{Fig:2}
\end{figure}


\begin{figure}[t!]
    \begin{center}
      \epsfig{file=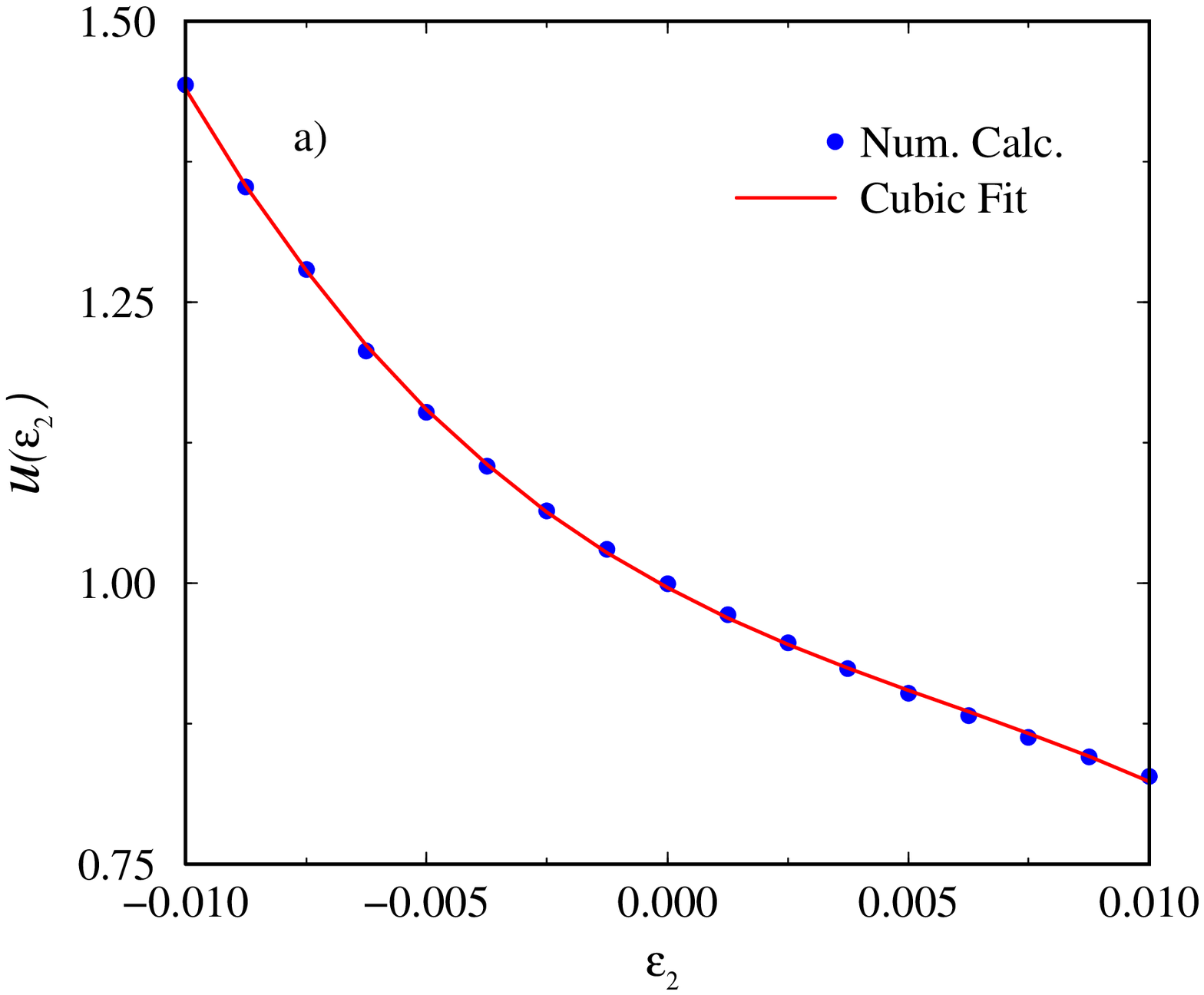, width=7.5cm,height=4.8cm} \\*[0.3cm]
      \epsfig{file=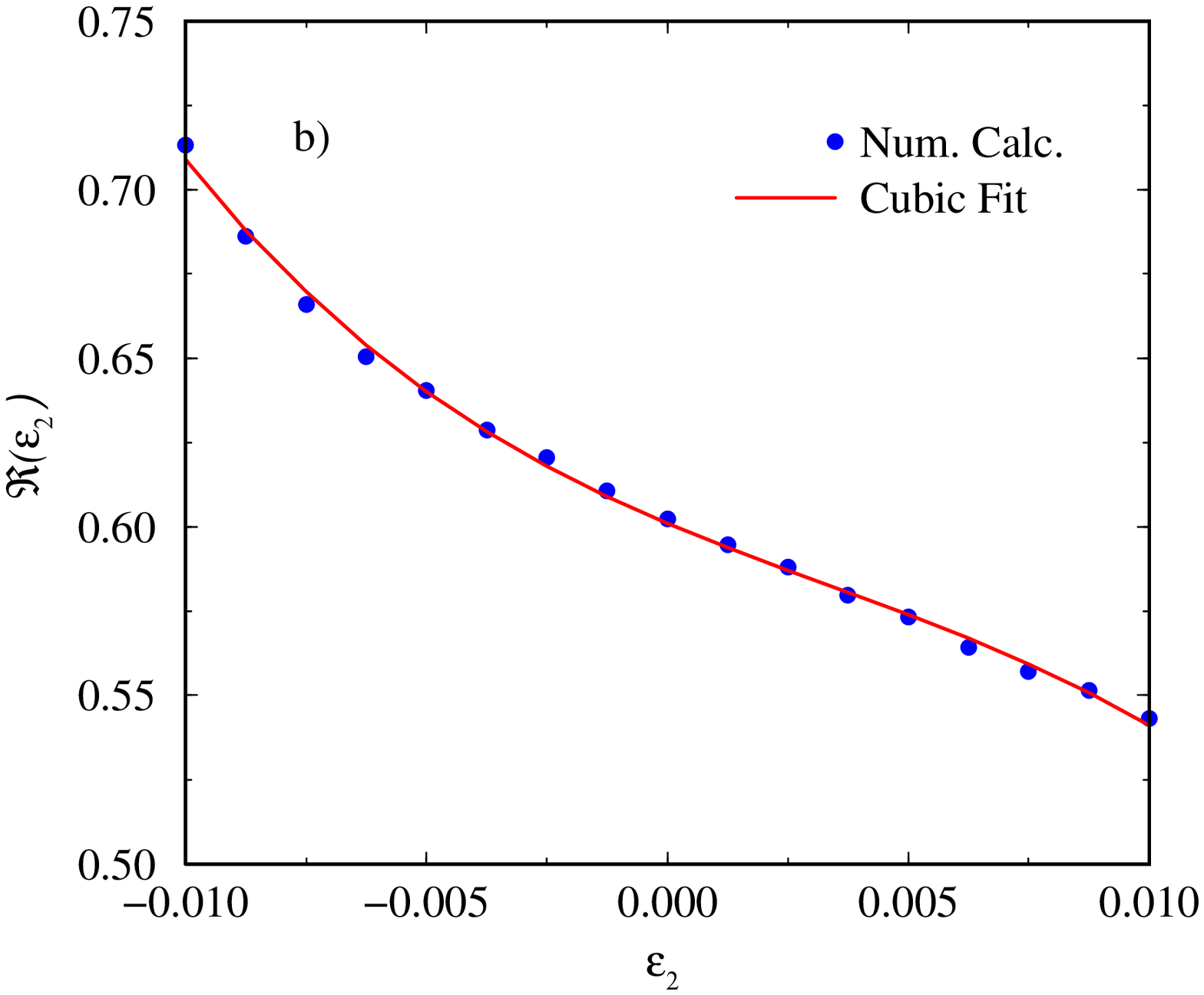, width=7.5cm,height=4.8cm}  \\*[0.3cm]
      \epsfig{file=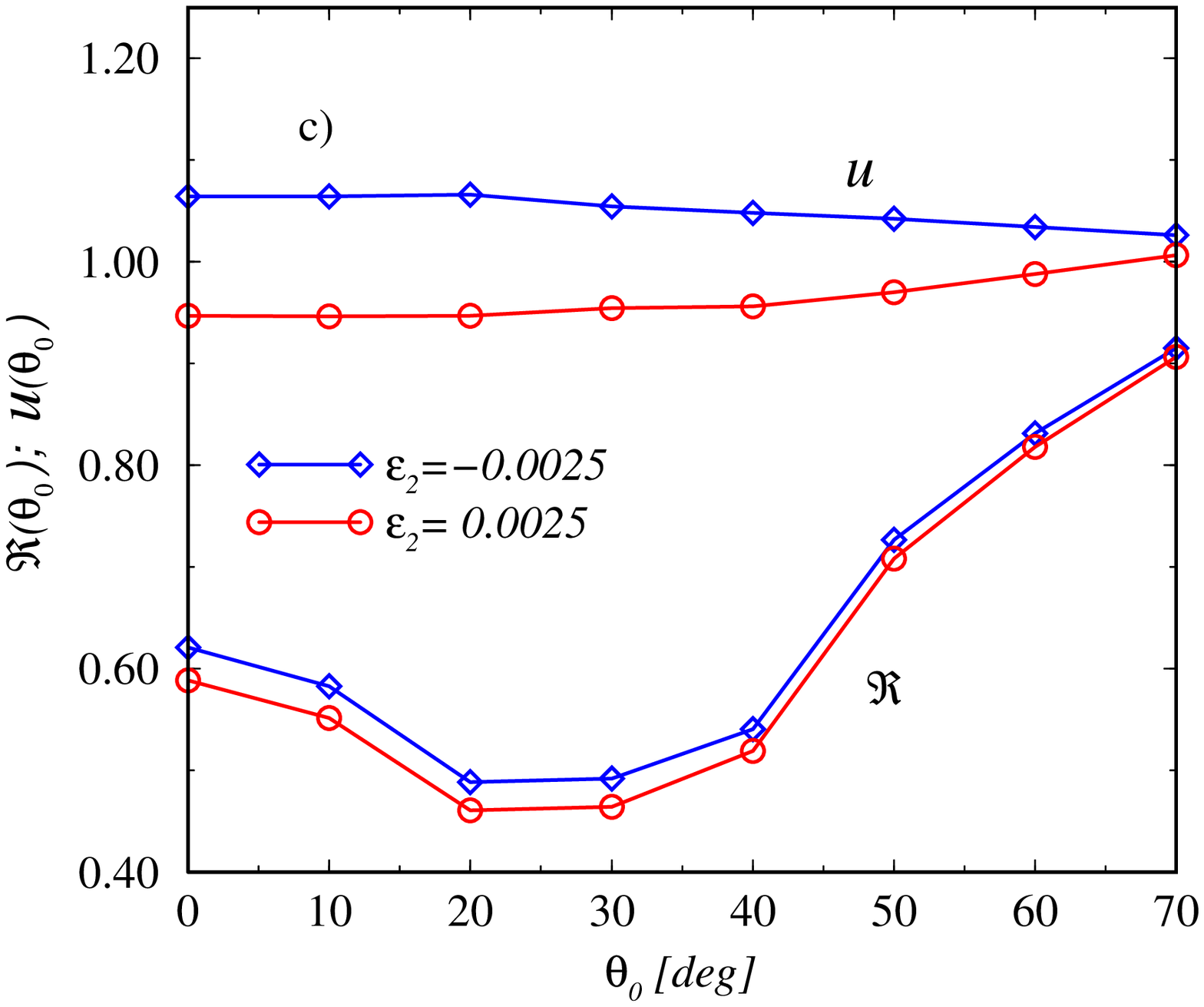, width=7.5cm,height=4.8cm}
    \end{center}
 \caption{The total scattered energy ${\cal
     U}$~(Eq.~\protect(\ref{Eq:total-energy})) and the reflectance
   ${\cal R}$(Eq.~\protect(\ref{Eq:reflectance})) as functions of the
   imaginary part of the dielectric
   constant~(Figs.~\protect\ref{Fig:3}a and 3b) and of the angle of
   incidence~(Fig.~\protect\ref{Fig:3}c). In
   Figs.~\protect\ref{Fig:3}a and b the angle of incidence was
   $\theta_0=0^\circ$, while in Fig.~\protect\ref{Fig:3}c the
   imaginary part of the dielectric constant was $\varepsilon_2=\pm
   0.0025$. The remaining parameters are as given in
   Fig.~\protect\ref{Fig:2}.}
    \label{Fig:3}
\end{figure}

To better quantify how the amplification or absorption depends on
$\varepsilon_2$, we have studied the total energy scattered by the
surface as well as its reflectance. These two quantities, denoted by
${\cal U}(\theta_0,\varepsilon_2)$ and ${\cal
  R}(\theta_0,\varepsilon_2)$ respectively, are related to the mean
differential reflection coefficient by Eqs.~(\ref{Eq:reflectance}) and
(\ref{Eq:total-energy}). The numerical results for these two
quantities for normal incidence are given in Figs.~3a and 3b. For
small values of $\varepsilon_2$ we find that these quantities are
linear in $\varepsilon_2$. However, when the absolute value of the
imaginary part of the dielectric constant increases, a deviation from
this behavior is observed. The numerical data in both cases are well
fitted by a cubic polynomial in $\varepsilon_2$.  In Fig.~3c we
present the numerical results for ${\cal U}$ and ${\cal R}$ as a
function of the angle of incidence $\theta_0$ for
$\varepsilon=\pm0.0025$. It is seen that ${\cal U}(\theta_0)$ is a
monotonically increasing or decreasing function of the angle of
incidence for positive and negative values of $\varepsilon_2$,
respectively, and the two curves for $\varepsilon_2=\pm 0.0025$ are
symmetric with respect to the line ${\cal U}(\theta_0)=1$. For
negative~(positive) $\varepsilon_2$ the total scattered energy is
larger~(smaller) than unity.  However, from the same graph it is
observed that ${\cal R}(\theta_0)$ is not a monotonic function of
$\varepsilon_2$. Instead it has a minimum in the vicinity of
$25^\circ$.  Below this value it is decreasing, while above it is
increasing. The reason for this behavior is due to the excitation of a
leaky guided wave supported by the scattering geometry~\cite{Tutov}.
The minimum in ${\cal R}(\theta_0)$ occurs for an angle of incidence
corresponding to the wave number of the leaky guided wave, and the
excitation of this mode will take away scattered energy from the
specular direction resulting in a minimum in ${\cal
  R}(\theta_0,\varepsilon_2)$ for this angle of incidence.


\begin{figure}[t!]
 \begin{center}
      \epsfig{file=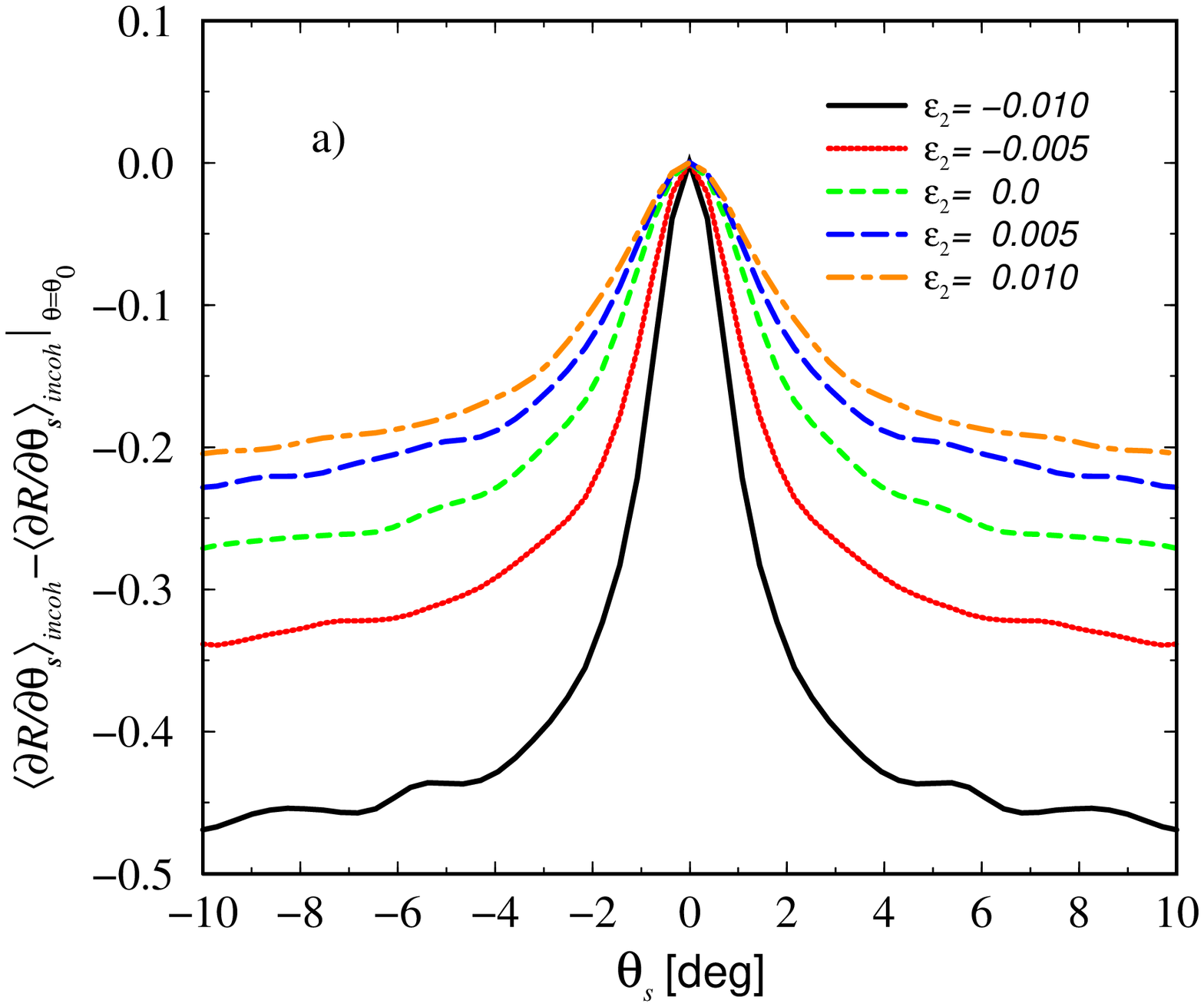, width=7.5cm,height=5cm}\\*[0.3cm]
      \epsfig{file=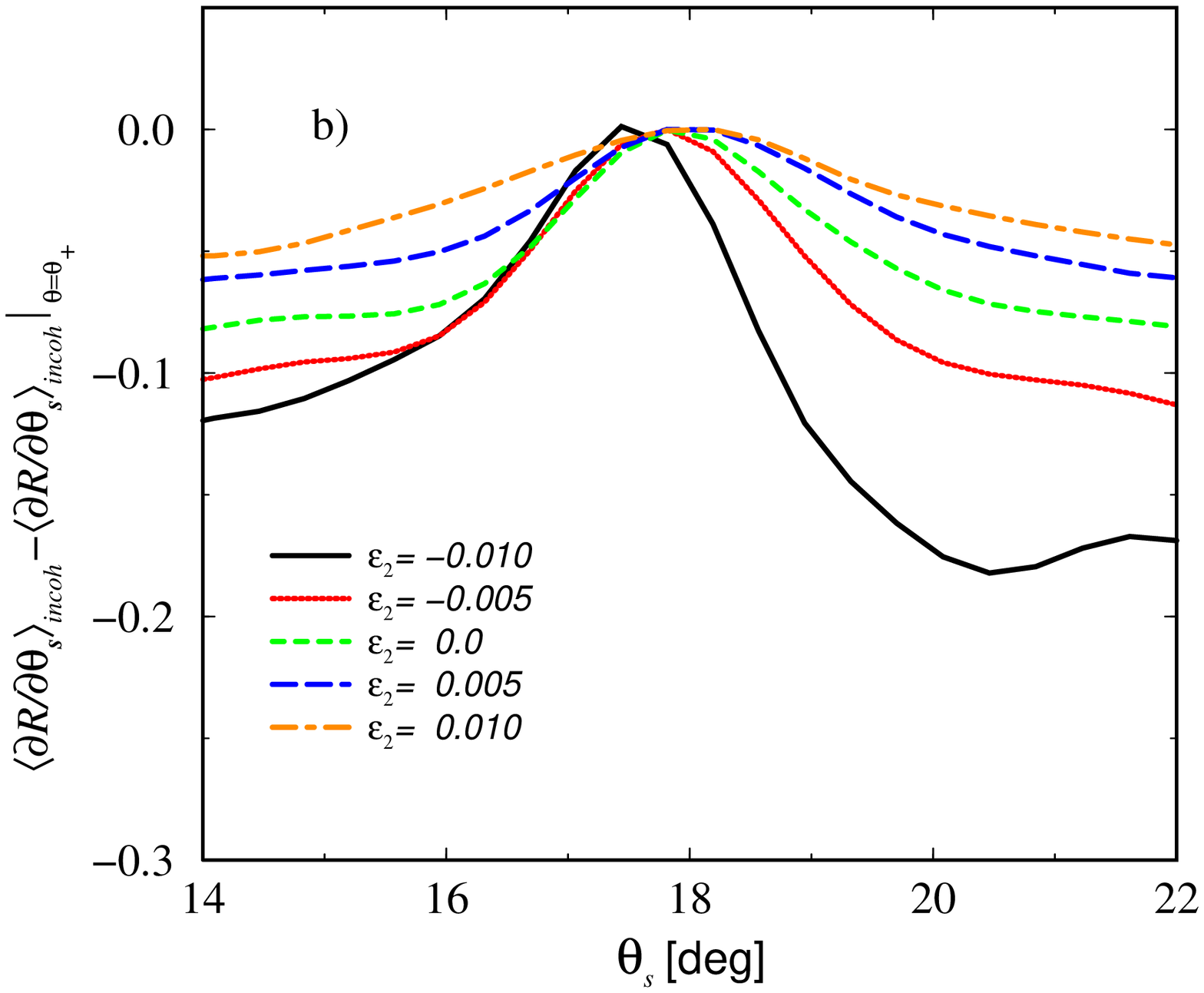,width=7.5cm,height=5cm}
    \end{center}
    \caption{Shifted plot for the mean DRC around the enhanced backscattering
      peak~(Fig.~\protect\ref{Fig:4}a) and satellite
      peaks~(Fig.~\protect\ref{Fig:4}b). The quantities that are
      plotted is $\left<\partial R /\partial \theta_s \right>_{{\rm
          incoh}}-\left.\left<\partial R /\partial \theta_s
        \right>_{{\rm incoh}}\right|_{\theta_s=\theta_o}$ for the
      backscattering peaks and $\left<\partial R /\partial \theta_s
      \right>_{{\rm incoh}}-\left.\left<\partial R /\partial \theta_s
        \right>_{{\rm incoh}}\right|_{\theta_s=\theta_+}$ for the
      satellite peaks, where $\theta_+$ is the (positive) angular
      position of the satellite peaks.}
    \label{Fig:4}
\end{figure}


From Fig.~2a it can be observed that the widths of both the
backscattering and satellite peaks, in contrast to their positions,
are sensitive to the value of the imaginary part of the dielectric
function. Since the scattering geometry supports (at least) two true
guided modes, the widths of these peaks are expected to grow with
$\varepsilon_2(\omega)$~\cite{Tutov}. This is much more apparent if we
shift, but not scale, the tops of the enhanced backscattering peaks to
the same height. We have done so by plotting $\left<\partial R
  /\partial \theta_s \right>_{{\rm incoh}}-\left.\left<\partial R
    /\partial \theta_s \right>_{{\rm
      incoh}}\right|_{\theta_s=\theta_o}$ as a function of the
scattering angle $\theta_s$ for various values of $\varepsilon_2$, and
the results are shown in Figs.~4 for the backscattering
peaks~(Fig.~4a) and the satellite peaks~(Fig.~4b). Figure~4a clearly
shows that the width of the enhanced backscattering peak increases as
the imaginary part of the dielectric constant increases. Or, in other
words, the enhanced backscattering peak becomes narrower and taller
when the amplification of the medium is increased. This behavior is in
qualitative agreement with the experimental results reported recently
by Gu and Peng~\cite{Gu}.  This finding can theoretically be
understood as follows: It can be shown that the enhanced
backscattering peak should have a Lorentzian form of total
width~\cite{Tutov}
\begin{eqnarray}
  \label{Eq:width}
  \Delta_T(\omega) &=& \Delta_\epsilon(\omega) + \Delta_{\rm sc}(\omega),
\end{eqnarray}
where $\Delta_\epsilon(\omega)$ is the contribution to the width from
the attenuation or amplification of the guided waves, while
$\Delta_{\rm sc}(\omega)$ is the broadening due to the scattering of such
waves by the surface roughness. Moreover, it can be shown that~\cite{Tutov} 
\begin{eqnarray}
  \label{Eq:Delta_eps}
  \Delta_\epsilon(\omega) \propto \varepsilon_2.
\end{eqnarray}
Depending on the geometrical and dielectric parameters of the film the
total width $\Delta_T(\omega)$ can be dominated by either
$\Delta_\epsilon(\omega)$ or $\Delta_{\rm sc}(\omega)$.  For the
parameters considered in this study, however, it is
expected~\cite{Tutov} that $\Delta_{\rm
  sc}(\omega)>|\Delta_\epsilon(\omega)|>0$.
Therefore  the width should increase
with increasing values of the imaginary part of the dielectric
constant.

We will now examine how the full width, $W(\varepsilon_2)$, of the
backscattering peak depends on $\varepsilon_2$. In Fig.~5 we
present $W(\varepsilon_2)$ vs.  $\varepsilon_2$ as obtained from the
numerical simulation results shown in Fig.~2. The width
of an  enhanced backscattering peak was defined as its full-width
at half maximum above the background at the position of the
 peak.  Here the background was defined to be located at
the minimum value of the $\left< \partial R/\partial
  \theta_s\right>_{incohr}$ between the backscattering and satellite
peaks. Even though the data in Fig.~5 are somewhat noisy a
linear dependence~(solid curve) on $\varepsilon_2$, as predicted by
Eq.~(\ref{Eq:Delta_eps}) is easily seen.

\begin{figure}
  \begin{center}
    \epsfig{file=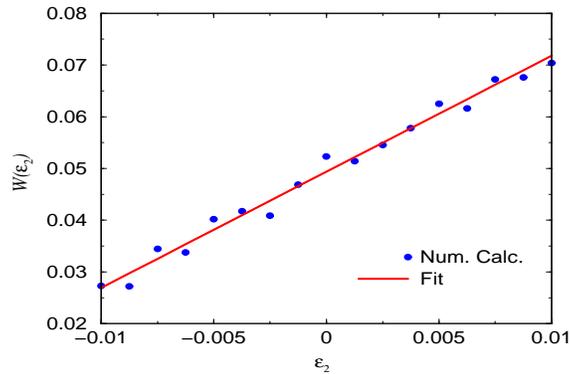, width=7.5cm,height=5cm} 
  \end{center}
  \caption{
    The full-width $W(\varepsilon_2)$ (filled dots) at half maximum
    above the background at its position of the backscattering peak as
    a function of the imaginary part $\varepsilon_2$ of the dielectric
    constant of the film as obtained from the numerical simulation
    results of Figs.~\protect\ref{Fig:2}. The solid line represents a
    linear fit in $\varepsilon_2$ to the numerical data.}
  \label{Fig:5}
\end{figure}

In Fig.~4b we present the same kind of plot as in
Fig.~4a, but now for a satellite peak. One sees the that the
width of the satellite peaks increases with increasing $\varepsilon_2$,
the same behavior found for the enhanced backscattering peak.
However, more interesting is the apparent change in the position of
the satellite peaks with the value of the imaginary part of the
dielectric constant.  To the precision of the numerical calculations,
the positions of the satellite peaks for an absorbing film
($\varepsilon_2>0$) seem to shift to larger scattering angles as
compared to the position of the satellite peaks when
$\varepsilon_2=0$. The opposite seems to hold true for an amplifying
film~($\varepsilon_2<0$).  There are two reasons for this behavior 
of the satellite peaks. First, 
the real part of the self--energy has a linear in   $\varepsilon_2$ 
contribution, thus, in the presence of surface roughness 
the values of the wavenumbers of 
the guided waves  acquire a linear in   $\varepsilon_2$ contribution.
The second reason is a strong dependence of the background
intensity  on the values of $\varepsilon_2$. The increase of the
background intensity also shifts the visual positions of the
satellite peaks to smaller scattering angles.
For the widths of the
satellite peaks, the quality of the numerical data, unfortunately, did
not allow us to obtain reliable results.

\section{Conclusions}

By numerical simulations we have studied light scattered from an
absorbing or amplifying dielectric film deposited on the planar
surface of a semi-infinite perfect conductor where the
vacuum-dielectric interface is randomly rough. It has been shown that
the reflectance, ${\cal R}(\theta_0,\varepsilon_2)$, as well as the
total scattered energy, ${\cal U}(\theta_0,\varepsilon_2)$, are
decreasing functions of the imaginary part of the dielectric function
of the film for a fixed angle of incidence.  Furthermore, it has been
demonstrated that ${\cal U}(\theta_0,\varepsilon_2)$ is a
monotonically increasing or decreasing function of the angle of
incidence for fixed positive and fixed negative values of the
imaginary part of the dielectric function, respectively.  However, for
the reflectance we find that ${\cal R}(\theta_0,\varepsilon_2)$ first
decreases to a minimum near $\theta_0=25^\circ$ and then 
increases.  This minimum is  a result of the
leaky guided wave supported by the scattering structure. Moreover, for
an amplifying surface both ${\cal R}(\theta_0,\varepsilon_2)$ and
${\cal U}(\theta_0,\varepsilon_2)$ are smaller then their absorbing
equivalents for all angles of incidence.

The width of the enhanced backscattering peaks as well as the
satellite peaks supported by the scattering system are found to increase
with increasing $\varepsilon_2$. While the location of the enhanced
scattering peaks seems to be unaffected by the value of the imaginary
part of the dielectric constant of the film, the corresponding
positions for the satellite peaks are found to be shifted towards
larger (smaller) scattering angles for positive (negative) values of
the imaginary part of the dielectric function, respectively.  Finally,
it is found that the width of the enhanced backscattering peak scales
linearly with $\varepsilon_2$.

\section*{Acknowledgments}
 
I.S. would like to thank the Research Council of Norway (Contract No.
32690/213) and Norsk Hydro ASA for financial support. The research of
I.S., A.A.M., T.A.L. was supported in part by Army Research Office
Grant DAAG 55-98-C-0034. This work has also received support from
the Research Council of Norway (Program for Supercomputing) through a
grant of computer time.


\stop 



\newpage
\noindent
\begin{figure}
    \caption{The scattering geometry considered in the present
      work.}
    \label{Fig:1}
\end{figure}

\begin{figure}
    \caption{The mean differential reflection coefficient for the
      incoherently scattered light for (a) $\varepsilon_2=0,\pm
      0.0025$ and (b) as a function of the same parameter. For both
      figures the angle of incidence was $\theta_0=0^\circ$, and the
      wavelength of the incident light was $\lambda=632.8\,{\rm nm}$.
      The dielectric function of the film of mean thickness $d=500 {\rm
        nm}$ was $\varepsilon(\omega) = 2.6896+i\varepsilon_2$, where
      $\varepsilon_2$ is as indicated in  the figure. The randomly
      rough surface had an rms roughness of $\delta=30 {\rm nm}$.
      The power spectrum was of the West-O'Donnell type
      defined by the parameters $k_-=0.86\, \omega/c$ and $k_+= 1.97\,
      \omega/c$.}
    \label{Fig:2}
\end{figure}

\begin{figure}
    \caption{The total scattered energy ${\cal
        U}$~(Eq.~\protect(\ref{Eq:total-energy})) and and the
      reflectance ${\cal R}$(Eq.~\protect(\ref{Eq:reflectance})) as
      functions of the imaginary part of the dielectric
      constant~(Figs.~\protect\ref{Fig:3}a and 3b) and of the angle of
      incidence~(Fig.~\protect\ref{Fig:3}c). In
      Figs.~\protect\ref{Fig:3}a and b the angle  of incidence was
      $\theta_0=0^\circ$, while in Fig.~\protect\ref{Fig:3}c the
      imaginary part of the dielectric constant was $\varepsilon_2=\pm
      0.0025$. The remaining parameters are as given in
      Fig.~\protect\ref{Fig:2}.}
    \label{Fig:3}
\end{figure}

\begin{figure}
    \caption{Shifted plot for the mean DRC around the enhanced backscattering
      peak~(Fig.~\protect\ref{Fig:4}a). The quantitie that is
      plotted is $\left<\partial R /\partial \theta_s \right>_{{\rm
          incoh}}-\left.\left<\partial R /\partial \theta_s
        \right>_{{\rm incoh}}\right|_{\theta_s=0}$.
      The full-width $W(\varepsilon_2)$ (filled dots)
      (Fig.~\protect\ref{Fig:4}b) at half 
      maximum above the background at its position of the backscattering 
      peak as a
      function of the imaginary part $\varepsilon_2$ of the dielectric
      constant of the film as obtained from the numerical simulation
      results of Figs.~\protect\ref{Fig:2}. The solid line represents
      a linear fit in $\varepsilon_2$ to the numerical data.}
    \label{Fig:4}
\end{figure}

\begin{figure}
    \caption{Shifted plot for the mean DRC around the satellite peak.
      The quantitie that is
      plotted is $\left<\partial R /\partial \theta_s \right>_{{\rm
          incoh}}-\left.\left<\partial R /\partial \theta_s
        \right>_{{\rm incoh}}\right|_{\theta_s=\theta_+}$,
       where $\theta_+$ is the (positive)
      angular position of the satellite peaks.}
    \label{Fig:5}
\end{figure}

\end{document}